\documentstyle[epsf,twoside,fleqn,espcrc2]{article}

\newcommand{\beq}{\begin{eqnarray}}
\newcommand{\eeq}{\end{eqnarray}}

\newcommand{\np}{Nucl.Phys.\ }
\newcommand{\pl}{Phys.Lett.\ }
\newcommand{\pr}{Phys.Rev.\ }
\newcommand{\asgen}{\alpha_s}

\newcommand{\MSB}{\overline{MS}}

\newcommand{\Gev}{{\rm GeV}}

\newcommand{\be}{\begin{equation}}
\newcommand{\ee}{\end{equation}}

\newcommand{\ewxy}[2]{\setlength{\epsfxsize}{#2}\epsfbox[-10 110 550
590]{#1}}

\def\np#1#2#3{Nucl.\ Phys.\ B#1 (19#3) #2}
\def\pl#1#2#3{Phys.\ Lett.\ #1B (19#3) #2}
\def\pr#1#2#3{Phys.\ Rev.\ D #1 (19#3) #2}

\def\zp#1#2#3{Zeit.\ Phys.\ C#1 (19#3) #2}

\newcommand{\AmS}{{\protect\the\textfont2
  A\kern-.1667em\lower.5ex\hbox{M}\kern-.125emS}}

\title{The running QCD coupling in the pre-asymptotic region}

\author{G. Burgio\address{Dipartimento di Fisica, Universit\`a di Parma
 and INFN, Gruppo Collegato di Parma, Parma, Italy}, 
        F. Di Renzo$^{\rm b}$, C. Parrinello\address{Dept. of 
Mathematical Sciences, University of Liverpool, 
Liverpool L69 3BX, U.K. \\
 (UKQCD Collaboration)}, C. Pittori\address{Institut de Physique 
Nucl\'eaire Th\'eorique, Universit\'e de Li\`ege au Sart 
Tilmann, B-4000 Li\`ege, Belgique}}

\begin{document}

\begin{abstract}
We study deviations from the perturbative asymptotic behaviour 
in the running QCD coupling by analysing non-perturbative measurements 
of $\asgen(p)$ at low momenta ($p \approx 2 \ \Gev$) as 
obtained from the lattice three-gluon vertex.
Our exploratory study provides some evidence for power corrections to 
the perturbative running proportional to $1/p^2$. 
\end{abstract}

\maketitle

\section{INTRODUCTION}
\label{sec:intro}
 
The standard procedure to parametrise non-perturbative QCD effects 
in terms of power corrections to perturbative results  
is based on the Operator Product Expansion (OPE). In this framework, 
the powers involved in the expansion are uniquely 
fixed by the symmetries and the dimension of the relevant operator 
product. 
The above picture has recently been challenged 
\cite{Akhoury,etc,Ceccobeppe}, when it was pointed out that  
 power corrections which are not {\it a priori} expected from OPE 
 may in fact appear in physical observables. 
 Such terms may arise from (UV-subleading) power corrections to 
$\asgen(p)$, corresponding to non-analytical contributions to the 
$\beta$-function. 
Clearly, the existence of OPE-independent 
power corrections, if demonstrated, 
 would have a major impact on our 
understanding of non-perturbative QCD effects and would affect 
QCD predictions for several processes.
For example, $\frac{\Lambda^2}{p^2}$ contributions 
may be relevant for the analysis of 
$\tau$ decays 
\cite{alt,Akhoury}. 

It would be highly desirable to 
develop a theoretical framework where the occurrence of these effects 
is demonstrated and estimates are obtained from first principles QCD 
calculations. The results in \cite{Lepage,Ceccobeppe} can be considered 
as a first step in this direction: some evidence for 
an unexpected $\frac{\Lambda^2}{Q^2}$ contribution to the gluon 
condensate was found by means of lattice calculations.
 
The aim of the present work is to test a method to detect  
the presence of power corrections in the running QCD coupling. 
Non-perturbative lattice estimates of the 
coupling at low momenta are compared with perturbative formulae. The 
final goal is to investigate the conjecture that OPE-independent power 
corrections to physical observables are linked to power terms in the 
running coupling.
Although at this stage our work is exploratory in nature and further 
simulations will be required to obtain a conclusive answer, 
our analysis provides some preliminary evidence for power corrections 
to $\asgen(p)$ for a particular definition of the coupling.


The paper is organised as follows: in Section \protect\ref{sec:relev} 
we briefly review some theoretical arguments in support of power 
corrections to $\asgen (p)$, illustrating the special role that may be 
played by $\frac{\Lambda^2}{p^2}$ terms. 
In Section \protect\ref{sec:lattice} 
we analyse the lattice data and present some preliminary 
evidence for power corrections. 
Finally, in Section \protect\ref{sec:conc} we draw our conclusions. 

\section{WHY POWER CORRECTIONS?}\protect\label{sec:relev}

Power corrections to $\asgen (p)$ can be shown to 
arise naturally in many physical schemes \cite{pino,maclep}.
Such corrections cannot be excluded {\em a priori} in any 
renormalisation scheme.
Clearly, the non-perturbative nature of such effects makes it very hard 
to assess their dependence on the renormalisation scheme.
 A term of order ${\Lambda^2}/{p^2}$ is a strong candidate for a 
power correction to $\asgen(p)$. To see why, consider 
 the interaction of two heavy quarks in the static limit  
and in the one-gluon-exchange approximation 
(for a more detailed discussion see \cite{zakEQ}). 
The static potential $V(r) $ can be written as
\begin{equation}
\protect\label{HQP}
V(r) \, \propto \, \alpha_s \ \int d^3p \,   
\frac{\exp^{i \vec{p} \cdot \vec{r}}}{|\vec{p}|^2}.
\end{equation} 
If one inserts in the above formula a running coupling of the form 
$\alpha_s(p^2) \approx {\Lambda^2}/{p^2}$, this results in a linearly 
confining potential. 
Similarly, consider the ``force" definition of the running 
coupling:
\begin{equation}
\alpha_{q\bar{q}}(Q) = \frac{3}{4} r^2 \frac{dV}{dr} \;\;\;\;\;
(Q = \frac{1}{r}),
\end{equation}
where again $V(r)$ represents the static interquark potential. 
A linear confinement term in $V(r)$ generates a 
$1/Q^2$ contribution to the coupling, whose order of magnitude is given 
by the string tension. 
This can be interpreted as a clue for the existence of a 
$\frac{\Lambda^2}{p^2}$ contribution, providing an estimate 
for its expected order of magnitude, at least in one (physically 
sound) scheme. 
Finally, power corrections to $\asgen(p)$ also 
emerge if one assumes that the 
singularities appearing in the perturbative formulae for the running 
coupling are ``removed" by non-perturbative effects \cite{RedBog}.

\section{ANALYSIS AND RESULTS}
\protect\label{sec:lattice}

We shall compare non-perturbative lattice data for $\asgen(p)$ with 
simple models where a power correction term is added to the perturbative 
formula at a given order. The first problem is the possible 
interplay between power corrections and our ignorance about higher 
orders of perturbation theory. In particular, for the scheme that we 
will consider, the three-loop coefficient of the $\beta$-function is not 
known. Knowledge of such a coefficient would allow a more reliable 
comparison of our estimates for the $\Lambda$ parameter in our scheme 
with lattice determinations  of $\Lambda$ in a different scheme, 
for which the three-loop result is available \cite{Lusch}.  
In fact, although matching the $\Lambda$ parameter between different 
schemes only requires a one-loop computation, the reliability of such a 
comparison rests on the assumption that the  value of $\Lambda$ in each 
scheme is fairly stable with respect to the inclusion of higher order 
terms in the definition of $\Lambda$.  
In practice, when working at two- or three-loop order, the value of 
$\Lambda$ is still quite sensitive to the order of the calculation. 
 Even within such limitations, we will argue that it is 
possible to estimate the impact of three-loop effects and that   
a description with power corrections seems relevant even at that order.

\subsection{Choice of the coupling}
 
We need to measure 
$\asgen (p)$ at low momenta (where power-like terms may be sizeable) and 
in a relatively wide momentum range. 
For this purpose, the best method is one where $\asgen (p)$ can be 
measured for several momentum values from a single Monte Carlo data set. 
One suitable method is the determination of 
the coupling from the renormalised lattice three-gluon vertex 
function \protect\cite{io,cpcp}.  
By varying the renormalisation scale $p$, one can determine $\asgen(p)$ 
for different momenta from a single simulation. Obviously the 
renormalisation scale must be chosen in a range such that finite volume 
effects and discretisation  errors are both under control. 
The numerical results for $\asgen(p)$ used in this work were 
obtained by applying such a method on a sample of 150 Monte Carlo 
configurations on a $16^4$ lattice at $\beta=6.0$. The calculation was 
performed in the Landau gauge.
For full details of the method we refer the reader to Ref. 
\protect\cite{cpcp}, where such results were first presented.
In order to detect violations of rotational invariance, different
combinations of lattice vectors have sometimes been used 
for a fixed value of $p^2$, which accounts for the graphical 
``splitting" of some data points.
 
\subsection{Two-loop analysis} 

At the two-loop level, we consider the following formula:
 
\begin{eqnarray}
\protect\label{2lp}
\alpha_s(p) \, = \, \frac{1}{b_0 \, \log(p^2/\Lambda_{2l}^2)} \, - \, 
\frac{b_1}{b_0} 
\frac{\log(\log(p^2/\Lambda_{2l}^2))}{(b_0 \, 
\log(p^2/\Lambda_{2l}^2))^2}
\, \nonumber \\ + \, c_{2l} \, \frac{\Lambda_{2l}^2}{p^2}
\qquad \qquad \qquad \qquad \qquad \qquad \end{eqnarray}

By fitting the data to (\protect\ref{2lp}) we obtain two 
estimates for ($\Lambda_{2l}$,$c_{2l}$), namely 
($0.84(1)$,$0.31(3)$) and ($0.73(1)$,$0.99(7)$), with comparable 
values for $\chi^2_{dof} \leq 1.8$. 
The momentum range for the fit corresponds to  $p \sim 1.8 - 3$ \Gev. 
We take the first set of values as our best estimate of the parameters 
as the corresponding value of $\Lambda_{2l}$ is close to 
what is obtained from a ``pure" two-loop fit, i.e. $\Lambda_{2l}$ is 
stable with respect to the introduction of power 
corrections. This choice will be supported also by 
independent considerations at the three-loop level. 
In summary, a two-loop description with power corrections based on 
(\protect\ref{2lp}) fits well the data in a consistent momentum range. 
Our best fit is shown in Figure 1.

\begin{figure}[htb]\protect\label{fig:2loopot}
\vspace{9pt}
\ewxy{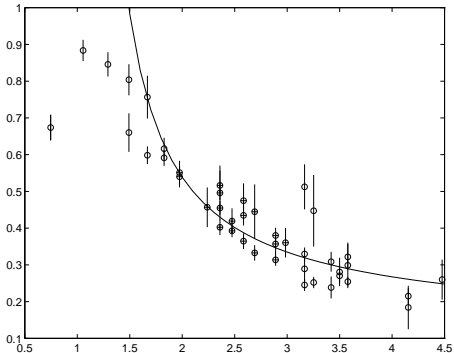}{70mm}
\caption{The best fit to (\protect\ref{2lp}).}
\end{figure}
 
\subsection{Three-loop analysis} 
 
A major obstacle for a three-loop analysis is our ignorance of  
the first non-universal  coefficient $b_2$ of the perturbative 
$\beta$-function.
In order to gain insight, we perform a two-parameter fit to 
the ``pure" three-loop formula, taking $\Lambda_{3l}$ and the unknown 
coefficient $b_2$ as the fitting parameters. We call 
$b_2^{eff}$ the fit estimate for $b_2$.
We obtain $\Lambda_{3l} = 
0.72(1)$, $b_2^{eff} = 1.3(1)$, with $\chi^2_{dof} \approx 1.8$ 
(see dashed curve in Fig. 2).
The momentum range where we obtain the best description of the 
data is $p \sim 2 - 3$ \Gev. Our result for 
$\Lambda_{3l}$ provides (via perturbative 
matching) an estimate for $\Lambda_{\MSB}$ which is in very good 
agreement with the estimate in \cite{Lusch}, which was obtained from 
the computation of the $\Lambda$ parameter in a completely different 
scheme. Although our estimate depends on the extra parameter 
$b_2^{eff}$, the agreement between the two results is remarkable. 

So far, the success of the ``pure" three-loop fit suggests that the 
power term in the two-loop formula merely provides an effective 
description of three-loop effects. However, it turns out that there 
is room for a power correction even at the three-loop level. To see 
this, we consider a three-loop formula with a power correction: 

\begin{eqnarray}
\protect\label{3lp}
\alpha_s(p) =  \frac{1}{b_0 \, L}  \, - \,  
\frac{b_1}{b_0} 
\frac{\log(L)}{(b_0 \, L)^2} \qquad \qquad \qquad \nonumber \\
+ \, \frac{1}{(b_0 \, L)^3} \, 
\left( \frac{b_2^{eff}}{b_0} + \frac{b_1^2}{b_0^2} 
(\log^2(L) - \log(L) + 1 ) \right) \nonumber \\
+ \, c_{3l} \, \frac{\Lambda_{3l}^2}{p^2}, \qquad \qquad 
\qquad \qquad \qquad \qquad 
\end{eqnarray}

\begin{figure}[htb]\protect\label{fig:3loopot}
\vspace{9pt}
\ewxy{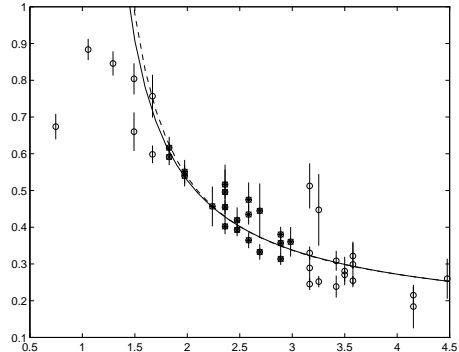}{70mm}
\caption{Fits to (\protect\ref{3lp})
(solid line) versus a pure three-loop fit (dashed line).}
\end{figure}
where $L = \log(p^2/\Lambda_{3l}^2)$ and 
 $b_2^{eff}$ is again to be determined from a fit.
Fitting the data to (\protect\ref{3lp}), we obtain
$\Lambda_{3l} = 0.72(1)$, $b_2^{eff} = 1.0(1)$ 
and $c_3 = 0.41(2)$, with $\chi^2_{dof} \approx 1.8$, in a momentum 
range $1.8 \, \Gev < p < 3 \, \Gev$ (see Fig. 2). 
We note that the value for $\Lambda_{3l}$ 
is fully consistent with the previous determination from the ``pure" 
three-loop description. The value for $b_2^{eff}$ is also reasonably 
stable with respect to the previous determination.
By comparing results from fits to (\protect\ref{2lp}) and 
(\protect\ref{3lp}), it emerges that 
\begin{equation}
c_2 \Lambda_{2l}^2 = 0.22(2) \, \Gev^2 \sim c_3 \Lambda_{3l}^2 = 0.21(2)
 \, \Gev^2.
\protect\label{eq:tuttotiene}
\end{equation}
In other words, the power terms providing the best fit to 
(\protect\ref{2lp}) and (\protect\ref{3lp}) are numerically the same, 
so that there seems to be no interplay between
the indetermination connected to the perturbative terms and the 
power correction term, within the precision of our data. 
We take this fact as an indication that a description in terms of power 
corrections is still relevant at the three-loop level.
Notice that the numerical value of the power correction
is comparable to the standard estimate for the string tension. 

One could object that at the two-loop level we had  chosen between 
two sets of values for ($\Lambda_{2l}$,$c_{2l}$), and that our choice 
is crucial for the validity of (\protect\ref{eq:tuttotiene}). 
 An {\it a posteriori} justification for our choice is obtained 
from the following test: we plot a few values for $\asgen(p)$ as 
generated by the ``pure'' three-loop formula 
for $\Lambda_{3l}= 0.72$ and $b_2 = 1.0$. Then, by fitting such 
points to the ``pure'' two-loop formula, one gets $\Lambda_{2l} 
\approx 0.84$, i.e. the value for which (\protect\ref{eq:tuttotiene}) 
holds. Again, the above test seems to confirm that perturbative and 
non-perturbative contributions do not mix in our formulae  
when upgrading from a two-loop to a three-loop description, thus 
suggesting that a genuine $\frac{\Lambda^2}{p^2}$ 
correction is present in  the data.  

\section{CONCLUSIONS}
\protect\label{sec:conc}

We have discussed an exploratory investigation of power corrections 
in the running QCD coupling $\asgen(p)$ by comparing non-perturbative 
lattice results with theoretical models. Some evidence was 
found for $1/p^2$ corrections, whose size would be consistent  
with what is suggested by simple arguments from the static potential.

Our results need to be further tested by the analysis of a larger data 
set and by a study of the dependence of the fit coefficient on the 
ultraviolet and infrared lattice cutoff. 
A very delicate issue is the assessment of the scheme dependendence of 
our results. In particular, we note that the definition of the coupling 
that we adopted is {\it a priori} gauge-dependent. This point will 
be the focus of our future work.
  
\section{ACKNOWLEDGEMENTS} 
We thank B. Alles, H. Panagopoulos and D. G. Richards for allowing us 
to use data files containing the results of Ref. \protect\cite{cpcp}.   
C. Parrinello acknowledges the support of PPARC
through an Advanced Fellowship.
C. Pittori thanks J. Cugnon and the ``Institut de 
Physique de l'Universit\'e de Li\`ege au Sart Tilman" 
and acknowledges the partial support of IISN.
We thank C. Michael for stimulating discussions.

\end{document}